\documentclass[a4paper,fleqn]{cas-sc}

\usepackage[authoryear,longnamesfirst]{natbib}
\usepackage{comment}
\newcommand{\sys}{StreamHD\xspace}
\usepackage{xspace}
\usepackage{algorithm}
\usepackage[noend]{algpseudocode}
\usepackage{comment}
\def\tsc#1{\csdef{#1}{\textsc{\lowercase{#1}}\xspace}}
\tsc{WGM}
\tsc{QE}

\begin{document}
\let\WriteBookmarks\relax
\def\floatpagepagefraction{1}
\def\textpagefraction{.001}

\shorttitle{\sys: Streaming FPGA Graph Classification}    

\shortauthors{Jebacyril Arockiaraj et al.}

\title[mode=title]{\sys: A Streaming FPGA Architecture for Graph Classification with Hyperdimensional Computing}

\author{Jebacyril Arockiaraj}[orcid=0009-0005-1926-852X]
\ead{arockiar@usc.edu}

\author{Dhruv Parikh}[orcid=0009-0000-6144-3858]
\ead{dhruvash@usc.edu}

\author{Viktor Prasanna}[orcid=0000-0002-1609-8589]
\ead{prasanna@usc.edu}

\affiliation{
  organization={University of Southern California},
  city={Los Angeles},
  state={California},
  country={USA}
}

\begin{abstract}
Real-time, energy-efficient inference on edge devices is essential for graph classification across a range of applications. Hyperdimensional Computing (HDC) is a brain-inspired computing paradigm that encodes input features into low-precision, high-dimensional vectors with simple element-wise operations, making it well-suited for resource-constrained edge platforms. Recent work enhances HDC accuracy for graph classification via Nyström kernel approximations. Existing FPGA implementations of Nyström-based HDC, however, primarily target small graph benchmarks and execute processing stages sequentially. This limits support for large graphs and increases end-to-end inference latency.

To address these challenges, we propose \textbf{\sys}, a streaming FPGA architecture for Nyström-based HDC graph classification at the edge that supports graphs exceeding on-chip storage capacity while reducing end-to-end inference latency. Implemented on an AMD Zynq UltraScale+ (ZCU104) FPGA, \sys achieves an average $3.1\times$ speedup over the non-streaming implementation. Compared with optimized CPU and GPU baselines, \sys achieves $14.3\times$ ($8.4\times$) speedup and $199.7\times$ ($273.8\times$) better energy efficiency, respectively.
\end{abstract}






\maketitle

\section{Introduction}
\label{sec:introduction}

Graphs are a natural representation for capturing relationships between entities in diverse domains such as bioinformatics~\cite{Borgwardt2005}, chemistry~\cite{molecule}, cybersecurity~\cite{Hu_Zhang_Shi_Zhou_Li_Qi_2019}, and social network analysis~\cite{10.1145/2783258.2783417}. Graphs exhibit irregular structure and variable size, making them difficult to process using conventional machine learning techniques. As a result, graph classification has emerged as a critical task and has received significant attention from the scientific community in recent years~\cite{errica2022faircomparisongraphneural}. Graph classification assigns a label to an entire graph, such as a molecule or biological structure, based on its topology and node attributes. There is a growing demand to bring such capabilities onto the edge for applications such as molecular toxicity screening and wearable biochemical sensing~\cite{rasool2025enhancing, GrapHD, NysHD}, which require real-time decisions under limited energy budgets~\cite{edge_intelligence_1, edge_intelligence_2}.

Graph Neural Networks (GNNs) are a class of models that extend deep learning to graphs through techniques such as iterative message passing and neighborhood aggregation, achieving high accuracy in tasks such as graph classification. However, they incur significant computational and energy costs, making them less suitable for resource-constrained edge devices~\cite{10.1145/3240765.3243473}.

Hyperdimensional Computing (HDC)~\cite{Kanerva2009HD} is a brain-inspired computing paradigm that represents information using low-precision vectors, often comprising thousands of dimensions. 
Unlike deep neural networks, HDC does not require complex iterative training and often generalizes well with minimal data and computation, enabling applications such as biosignal analysis and cybersecurity~\cite{static-encoding-1-programmable-hdc, static-encoding-2-seizure-det, static-encoding-3-colal-learning-secure-hdc, static-encoding-6-genie-hd, static-encoding-7-efficient-human-activity-recognition-hdc-imani, DistriHD, static-encoding-9-hdc-framework-image-descriptors-cvpr, static-encoding-11-imani-scalable-edge-hdc}. The lightweight, low complexity operations make HDC a compelling alternative to deep learning for low-power, real-time, and edge inference~\cite{10.1145/3489517.3530669, 10.1145/3495243.3558757}.  

A key step in HDC is feature encoding, which converts raw input features into hypervectors (HVs). 
Early graph-focused efforts~\cite{GrapHD} primarily captured graph topology and overlooked node attributes, limiting expressiveness. Recent work~\cite{NysHD} addresses this by enabling kernel functions in the encoding stage, allowing HDC to capture complex, nonlinear graph structures. 
During inference, given an input graph, the model performs multi-hop propagation, generates integer codes for each node, hashes them into the codebook, and updates the query graph histogram. The histogram is then compared against landmark histograms to form a kernel score, which is projected via the Nyström projection matrix into an HV for final classification.

By incorporating kernel methods into HDC encoding~\cite{NysHD}, the accuracy gap of HDC with GNN-based models for graph classification is significantly reduced. 
Existing work~\cite{HyperX} designs an end-to-end FPGA accelerator for Nyström-based HDC graph classification, addressing key computational and memory bottlenecks for relatively small graph datasets. However, large graphs arise in practical graph classification applications, such as malware analysis in IoT~\cite{Malnet,wu2023iot} and robotic object recognition~\cite{neumann2013graph}. Such graphs can exceed the limited on-chip storage capacity of FPGAs, requiring a scalable execution architecture. Moreover, histogram generation in Nyström-based HDC introduces a synchronization barrier between processing stages, preventing pipelined execution and increasing end-to-end inference latency.

To address these limitations, we propose \sys{}, a streaming FPGA architecture for Nyström-based HDC graph classification. \sys{} supports large graphs by processing the feature and adjacency matrices as row blocks streamed from off-chip memory. To reduce inference latency, \sys{} restructures the processing pipeline by fusing histogram update with kernel similarity computation in a dataflow architecture, eliminating the intermediate synchronization barrier and enabling pipelined execution.

The key contributions of this work are:

\begin{itemize}
\item We develop an FPGA architecture that scales Nyström-based HDC graph classification to large graphs exceeding on-chip storage capacity.

\item We introduce a streaming dataflow architecture that removes stage-wise synchronization barriers, enables pipelined execution within and across hops, and achieves an average $3.1\times$ speedup over the non-streaming implementation.

\item We evaluate \sys{} across small- and large-graph datasets. Compared with optimized CPU and GPU baselines, \sys{} achieves $14.3\times$ ($8.4\times$) speedup and $199.7\times$ ($273.8\times$) better energy efficiency, respectively.

\end{itemize}

This article extends the conference version of this work~\citep{HyperX}, which buffers each query graph on chip and executes the processing stages of each hop sequentially. Relative to that paper, this article (i) streams the feature and adjacency data of large graphs from DDR, supporting graphs that exceed on-chip memory; (ii) fuses histogram update and kernel similarity computation into a single engine with window accumulation; (iii) introduces a FIFO-based dataflow architecture with intra-hop and inter-hop pipelining; and (iv) extends the evaluation to three datasets with larger graphs, MalNet-Tiny, REDDIT-Multi-5K, and FIRSTMM\_DB.

\section{Background}
\label{sec:background}

\subsection{Hyperdimensional Computing}
\label{subsec:hdc}

Hyperdimensional Computing (HDC) is a brain-inspired paradigm that represents information as high-dimensional vectors, or \emph{hypervectors} (HVs), typically with dimensionality $d \sim 10^4$~\citep{Kanerva2009HD,Kleyko2023Survey}. HVs are low-precision; in this work, we use bipolar HVs $\mathbf{h} \in \{-1,+1\}^d$. Learning in HDC relies on a small set of inexpensive, element-wise, and noise-robust operations that are well suited for hardware acceleration~\citep{Imani2019BRIC,10.1145/3489517.3530669}:
\begin{itemize}
    \item \textbf{Bundling} ($\oplus$): element-wise addition followed by thresholding, e.g., $\mathbf{h} = \mathrm{sign}(\mathbf{h}_1 + \mathbf{h}_2)$, which combines HVs while preserving their similarity to the inputs.
    \item \textbf{Binding} ($\otimes$): element-wise multiplication, $\mathbf{h} = \mathbf{h}_1 \odot \mathbf{h}_2$, which produces an HV dissimilar to both inputs and is used to encode associations~\citep{Plate1995HRR}.
    \item \textbf{Permutation} ($\rho$): a cyclic shift, $\rho^i(\mathbf{h}) = [h_{(j+i)\bmod d}]_{j=0}^{d-1}$, which encodes position or sequence information~\citep{Joshi2017LanguageGeometry}.
\end{itemize}
An HDC classifier stores one \emph{class prototype} $\mathbf{g}_c \in \{-1,+1\}^d$ per class, obtained by bundling the encoded training samples corresponding to class $c$. The class prototypes can be further refined via a few retraining epochs~\citep{HernandezCano2021OnlineHD}. During inference, the query HV is compared against all prototypes, and the class with the highest similarity is predicted. Single-pass training, low-precision arithmetic, and bit-level parallelism make HDC attractive for edge-deployed learning systems~\citep{Salamat2019F5HD,Kang2022XcelHD,survey-edge}. The accuracy of an HDC model, however, depends largely on its \emph{encoder}, the function that maps raw inputs to HVs.

\subsection{Nyström-Based HDC Encoding}
\label{subsec:nystrom}

Conventional HDC encoders rely on random projections~\citep{Rahimi2007RFF,Thomas2021TheoreticalHDC,GrapHD}, which approximate only simple, typically shift-invariant, similarity functions and cannot capture the structured similarity of inputs such as graphs. NysHD~\citep{NysHD} addresses this limitation by incorporating kernel methods into HDC encoding through the Nyström method~\citep{Williams2000Nystrom,Kumar2012Sampling}, which turns any positive semi-definite kernel $K(\cdot,\cdot)$ into an explicit HDC encoder. For graph classification, it achieves state-of-the-art accuracy among HDC-based approaches and substantially narrows the accuracy gap to GNNs. Our prior work~\citep{HyperX} therefore adopts this algorithm and augments it with diversity-aware landmark selection (Section~\ref{subsec:dpp}). \sys{} accelerates the same algorithm, which we refer to as \emph{Nyström-HDC} and summarize in the remainder of this section.

The Nyström method approximates a kernel from its values on a small set of reference samples, called \emph{landmarks}. Let $\mathcal{Z}=\{z_1,\dots,z_s\}$ be $s \ll n$ training samples selected as landmarks from a training set of size $n$. Let $\mathbf{K}_{\mathcal{Z}}\in\mathbb{R}^{s\times s}$ denote the landmark kernel matrix with $(\mathbf{K}_{\mathcal{Z}})_{ij}=K(z_i,z_j)$, and let $\mathbf{K}_{n\mathcal{Z}}\in\mathbb{R}^{n\times s}$ collect the kernel values between all training samples and the landmarks. The Nyström method approximates the full $n\times n$ kernel matrix $\mathbf{K}$ as
\begin{equation}
\hat{\mathbf{K}} = \mathbf{K}_{n\mathcal{Z}}\,\mathbf{K}_{\mathcal{Z}}^{+}\,\mathbf{K}_{n\mathcal{Z}}^{\top},
\label{eq:nystrom}
\end{equation}
where $\mathbf{K}_{\mathcal{Z}}^{+}$ is the pseudoinverse of $\mathbf{K}_{\mathcal{Z}}$. This low-rank form can be factored into explicit features. With the eigendecomposition $\mathbf{K}_{\mathcal{Z}}=\mathbf{Q}\boldsymbol{\Lambda}\mathbf{Q}^{\top}$, the pseudoinverse is $\mathbf{K}_{\mathcal{Z}}^{+}=\mathbf{Q}\boldsymbol{\Lambda}^{-1}\mathbf{Q}^{\top}$, so $\hat{\mathbf{K}}=\boldsymbol{\Phi}\boldsymbol{\Phi}^{\top}$ with $\boldsymbol{\Phi}=\mathbf{K}_{n\mathcal{Z}}\mathbf{Q}\boldsymbol{\Lambda}^{-1/2}$. Each row of $\boldsymbol{\Phi}$ is an $s$-dimensional embedding of one training sample, and the same map applies to any input $x$: $\boldsymbol{\phi}_{\mathrm{nys}}(x)=\boldsymbol{\Lambda}^{-1/2}\mathbf{Q}^{\top}\mathbf{C}(x)$, where
\begin{equation}
\mathbf{C}(x)=\big[K(x,z_1),\,\dots,\,K(x,z_s)\big]^{\top}\in\mathbb{R}^{s}
\label{eq:kernel-vector}
\end{equation}
is the \emph{kernel similarity vector} between $x$ and the landmarks. Unlike the implicit feature map of $K$, $\boldsymbol{\phi}_{\mathrm{nys}}$ is explicit and requires only $s$ kernel evaluations per input, and its inner products reproduce the approximate kernel, i.e., $\boldsymbol{\phi}_{\mathrm{nys}}(x_i)^{\top}\boldsymbol{\phi}_{\mathrm{nys}}(x_j)=\hat{\mathbf{K}}_{ij}$.

Because $\boldsymbol{\phi}_{\mathrm{nys}}(x)$ is real-valued, Nyström-HDC maps it to a bipolar HV using a random hyperplane projection $\mathbf{P}_{\mathrm{rp}}\in\mathbb{R}^{d\times s}$, whose rows are sampled uniformly from the unit sphere, followed by sign thresholding~\citep{Charikar2002SimEst}. Folding both projections into a single \emph{Nyström projection matrix}
\begin{equation}
\mathbf{P}_{\mathrm{nys}}=\mathbf{P}_{\mathrm{rp}}\,\boldsymbol{\Lambda}^{-1/2}\mathbf{Q}^{\top}\in\mathbb{R}^{d\times s}
\label{eq:pnys}
\end{equation}
yields the encoder $\mathbf{h}=\mathrm{sign}\big(\mathbf{P}_{\mathrm{nys}}\mathbf{C}(x)\big)$. In expectation over $\mathbf{P}_{\mathrm{rp}}$, the (scaled) inner product of two encoded HVs matches the normalized approximate kernel $\hat{\mathbf{K}}$ to first order~\citep{NysHD}. HV similarity therefore preserves the kernel similarity on which HDC classification relies.

\subsection{Propagation Kernel for Graph Classification}
\label{subsec:propagation-kernel}

For graph classification, Nyström-HDC instantiates $K$ with the propagation kernel~\citep{Neumann2016Propagation}. Unlike Weisfeiler--Lehman kernels~\citep{Shervashidze2011WeisfeilerLehman}, which refine discrete node labels, the propagation kernel supports continuous node attributes by hashing them with locality sensitive hashing (LSH). It measures graph similarity by iteratively propagating node features through the graph and comparing histograms of the propagated representations.

Let $\mathbf{A}_x\in\{0,1\}^{N\times N}$ denote the adjacency matrix of a graph $G_x$ with $N$ nodes, $\mathbf{F}_x\in\mathbb{R}^{N\times f}$ its node feature matrix, and $H$ the number of propagation hops. Starting from $\mathbf{M}^{(0)}=\mathbf{F}_x$, node features are propagated as $\mathbf{M}^{(t+1)}=\mathbf{A}_x\mathbf{M}^{(t)}$. At each hop $t$, the feature vector $\mathbf{m}^{(t,i)}\in\mathbb{R}^{f}$ of node $i$ (the $i$-th row of $\mathbf{M}^{(t)}$) is mapped to an integer code using LSH:
\begin{equation}
c^{(t,i)}=\left\lfloor\frac{\big(\mathbf{m}^{(t,i)}\big)^{\top}\mathbf{u}^{(t)}+b^{(t)}}{w}\right\rfloor ,
\label{eq:lsh}
\end{equation}
where $\mathbf{u}^{(t)}\in\mathbb{R}^{f}$ is a random projection vector, $b^{(t)}\in\mathbb{R}$ a scalar offset, and $w>0$ a quantization width shared across hops. Each code is then looked up in a hop-specific \emph{codebook} $\mathcal{B}^{(t)}$, the vocabulary of codes observed on the landmark graphs during training. If $c^{(t,i)}\in\mathcal{B}^{(t)}$, the corresponding bin of the hop histogram $\mathbf{h}^{(t)}\in\mathbb{Z}^{|\mathcal{B}^{(t)}|}$ is incremented; otherwise, no update is made. The kernel between two graphs sums the histogram inner products over all hops:
\begin{equation}
K(G_x,G_z)=\sum_{t=0}^{H-1}\big(\mathbf{h}_x^{(t)}\big)^{\top}\mathbf{h}_z^{(t)} .
\label{eq:propagation-kernel}
\end{equation}
Stacking the hop-$t$ histograms of all landmarks as rows of the \emph{landmark histogram matrix} $\mathbf{H}^{(t)}\in\mathbb{R}^{s\times|\mathcal{B}^{(t)}|}$, the kernel similarity vector of Eq.~\eqref{eq:kernel-vector} becomes a sum of hop-wise similarity vectors:
\begin{equation}
\mathbf{C}=\sum_{t=0}^{H-1}\mathbf{v}^{(t)},\qquad \mathbf{v}^{(t)}=\mathbf{H}^{(t)}\mathbf{h}^{(t)}\in\mathbb{R}^{s}.
\label{eq:hop-similarity}
\end{equation}
Since each landmark graph populates only a small fraction of the codebook bins, $\mathbf{H}^{(t)}$ is sparse. Finally, the LSH code in Eq.~\eqref{eq:lsh} depends on $\mathbf{M}^{(t)}$ only through the projection $\mathbf{M}^{(t)}\mathbf{u}^{(t)}=\mathbf{A}_x^{t}\,\mathbf{F}_x\mathbf{u}^{(t)}$. The codes of hop $t$ can therefore be obtained by first projecting $\mathbf{F}_x$ onto $\mathbf{u}^{(t)}$ and then propagating the resulting length-$N$ vector $t$ times, without materializing the $N\times f$ matrix $\mathbf{M}^{(t)}$.

\subsection{Diversity-Aware Landmark Selection}
\label{subsec:dpp}

The landmark set governs both the approximation quality and the inference cost of Nyström-HDC: every query graph is compared against all $s$ landmarks, and the size of $\mathbf{P}_{\mathrm{nys}}\in\mathbb{R}^{d\times s}$ grows linearly with $s$. The original algorithm selects landmarks by uniform sampling~\citep{NysHD}, which does not guarantee diversity. Graphs with similar structure and node features may be selected repeatedly, increasing the projection cost while biasing the landmark set toward common patterns. In our prior work~\citep{HyperX}, we therefore introduced a hybrid landmark selection strategy that combines uniform sampling with a determinantal point process (DPP). DPPs~\citep{Kulesza_2012} favor subsets with high pairwise dissimilarity, but sampling from a DPP requires the full kernel matrix over the candidates and its eigendecomposition. Unlike DPP-based Nyström sampling for Euclidean data~\citep{fastdpp}, where kernel evaluations are inexpensive, every entry of this matrix is a propagation-kernel evaluation, so applying a DPP over all training graphs is impractical. The hybrid strategy therefore first draws a candidate pool by uniform sampling, of size $\max(300,\,0.02n)$ as in~\cite{NysHD}, then computes the propagation-kernel matrix over the pool and applies DPP sampling to select a smaller, diverse set of $s$ landmarks. Selection runs once, offline, during training. Fewer and more diverse landmarks shrink $\mathbf{P}_{\mathrm{nys}}$ and the landmark-similarity cost while improving accuracy. \sys{} adopts this landmark selection.

\subsection{End-to-End Nyström-HDC Inference}
\label{subsec:end-to-end-inf}

Algorithm~\ref{alg:end-to-end} summarizes end-to-end inference for a query graph $G_x=(\mathbf{A}_x,\mathbf{F}_x)$. The trained model consists of the hop-specific codebooks $\{\mathcal{B}^{(t)}\}$, landmark histogram matrices $\{\mathbf{H}^{(t)}\}$, LSH parameters $\{(\mathbf{u}^{(t)},b^{(t)})\}$ with width $w$, the Nyström projection matrix $\mathbf{P}_{\mathrm{nys}}$, and the class-prototype matrix $\mathbf{G}=[\mathbf{g}_1;\dots;\mathbf{g}_C]\in\{-1,+1\}^{C\times d}$. At each hop, node features are propagated, node codes are computed and binned into the hop histogram $\mathbf{h}^{(t)}$, and the histogram is compared against all landmark histograms through a single matrix--vector product to yield $\mathbf{v}^{(t)}$. Accumulating $\mathbf{v}^{(t)}$ over $H$ hops produces $\mathbf{C}$, which is encoded into an HV and matched against the class prototypes.

\begin{algorithm}
\small
\caption{End-to-End Inference for Nyström-HDC}
\label{alg:end-to-end}
\begin{algorithmic}[1]
\Require $\mathbf{A}_x\!\in\!\{0,1\}^{N\times N}$; $\mathbf{F}_x\!\in\!\mathbb{R}^{N\times f}$; hops $H$; codebooks $\{\mathcal{B}^{(t)}\}$; landmark histograms $\{\mathbf{H}^{(t)}\!\in\!\mathbb{R}^{s\times|\mathcal{B}^{(t)}|}\}$; LSH params $\{(\mathbf{u}^{(t)},b^{(t)})\}$ and width $w$; Nyström projection $\mathbf{P}_{\mathrm{nys}}\!\in\!\mathbb{R}^{d\times s}$; class prototypes $\mathbf{G}\!\in\!\{-1,+1\}^{C\times d}$
\Ensure Query HV $\mathbf{h}\!\in\!\{-1,+1\}^{d}$ and predicted class $\hat{y}$
\State $\mathbf{M}\!\gets\!\mathbf{F}_x$                                   \Comment{node features}
\State $\mathbf{C}\!\gets\!\mathbf{0}\in\mathbb{R}^{s}$                     \Comment{similarity accumulator}
\For{$t=0$ \textbf{to} $H-1$}
  \State $\mathbf{c}\!\gets\!\Big\lfloor(\mathbf{M}\mathbf{u}^{(t)}+b^{(t)}\mathbf{1}_N)/w\Big\rfloor\in\mathbb{Z}^{N}$ \Comment{LSH codes}
  \State $\mathbf{h}^{(t)}\!\gets\!\mathbf{0}\in\mathbb{Z}^{|\mathcal{B}^{(t)}|}$                                         \Comment{histogram}
  \For{$v=1$ \textbf{to} $N$}
    \State $c\!\gets\!\mathbf{c}[v]$
    \State \textbf{if} $c\!\in\!\mathcal{B}^{(t)}$ \textbf{then} $j\!\gets\!\textproc{Index}(\mathcal{B}^{(t)},c)$; $\mathbf{h}^{(t)}[j]\!\gets\!\mathbf{h}^{(t)}[j]+1$
  \EndFor
  \State $\mathbf{v}^{(t)}\!\gets\!\mathbf{H}^{(t)}\mathbf{h}^{(t)}\in\mathbb{R}^{s}$                 \Comment{hop similarity}
  \State $\mathbf{C}\!\gets\!\mathbf{C}+\mathbf{v}^{(t)}$ \Comment{accumulating into kernel similarity vector}
  \If{$t<H-1$} \State $\mathbf{M}\!\gets\!\mathbf{A}_x\mathbf{M}$ \Comment{propagate} \EndIf
\EndFor
\State $\mathbf{y}\!\gets\!\mathbf{P}_{\mathrm{nys}}\mathbf{C}\in\mathbb{R}^{d}$; \ \ $\mathbf{h}\!\gets\!\mathrm{sign}(\mathbf{y})$ \Comment{encoding}
\State $\hat{y}\!\gets\!\arg\max_{c\in\{1,\dots,C\}}\mathrm{sim}(\mathbf{h},\mathbf{g}_c)$ \Comment{similarity against class prototypes}
\State \Return $\mathbf{h}$ and $\hat{y}$
\end{algorithmic}
\end{algorithm}

Per query graph, Algorithm~\ref{alg:end-to-end} performs $H$ hops of propagation, codebook lookup, and landmark similarity, whose cost grows with $\mathrm{nnz}(\mathbf{A}_x)$ and the number of nonzeros in $\mathbf{H}^{(t)}$, followed by a single Nyström projection with $O(sd)$ operations and prototype matching with $O(Cd)$ operations. Table~\ref{tab:mem-complexity} summarizes the storage of inputs, intermediates, and model parameters. The first three rows scale with the size of the query graph, whereas the remaining rows are fixed by the trained model.

\begin{table}[t]
\centering
\caption{Storage of inputs, intermediates, and model parameters in end-to-end Nyström-HDC inference. $\mathbf{A}_x$ is stored in CSR format; $b_{(\cdot)}$ is the element bit width, and $b_{\mathcal{B}^{(t)}}$ covers a code and its histogram index.}
\label{tab:mem-complexity}
\small
\begin{tabular}{@{}lc@{}}
\toprule
\textbf{Component} & \textbf{Memory} \\ \midrule
Adjacency matrix $\mathbf{A}_x$ (CSR) & $(N{+}1)\,b_{\mathrm{ptr}}+\mathrm{nnz}(\mathbf{A}_x)\,(b_{\mathrm{idx}}+b_A)$ \\
Node features $\mathbf{F}_x$ & $N f\, b_F$ \\
Propagated features $\mathbf{M}$ & $N f\, b_F$ \\
Codebooks $\{\mathcal{B}^{(t)}\}$ & $\sum_{t} |\mathcal{B}^{(t)}|\, b_{\mathcal{B}^{(t)}}$ \\
Landmark histograms $\{\mathbf{H}^{(t)}\}$ & $\sum_{t} \mathrm{nnz}(\mathbf{H}^{(t)})\, b_H$ \\
Nyström projection $\mathbf{P}_{\mathrm{nys}}$ & $d s\, b_P$ \\
Class prototypes $\mathbf{G}$ & $C d\, b_G$ \\
\bottomrule
\end{tabular}
\end{table}

\paragraph{Problem definition.}
Given a labeled graph dataset $\mathcal{D}=\{(G_i,y_i)\}_{i=1}^{n}$ with $G_i=(\mathbf{A}_i,\mathbf{F}_i)$ and labels $y_i\in\{1,\dots,C\}$, graph classification seeks a mapping $\mathcal{F}:(\mathbf{A},\mathbf{F})\mapsto\{1,\dots,C\}$. We target \emph{single-graph, real-time} Nyström-HDC inference on resource-constrained edge FPGAs, where each query graph is classified independently on arrival (batch size one). For each query graph, we measure end-to-end latency (ms), from input arrival to prediction, and energy per graph (mJ). The goal is to minimize both metrics within the on-chip memory (BRAM/URAM), compute fabric (LUTs, DSPs), and external memory bandwidth (DDR) of the device. Unlike our prior work~\citep{HyperX}, we do not assume that a query graph fits on chip: the feature matrix $\mathbf{F}_x$ and the adjacency matrix $\mathbf{A}_x$ may exceed on-chip storage capacity.

\section{Related Work}
\label{sec:related}

\paragraph{HDC for graph learning.}
GraphHD~\citep{GrapHD} first applied HDC to graph classification, encoding graph topology through node centrality but ignoring node attributes. NysHD~\citep{NysHD} incorporates node attributes through the propagation kernel and the Nyström method, at the cost of kernel evaluations and a large projection matrix. Other HDC works target node classification and link prediction~\citep{RelHD,RelHDx,HDGL} or knowledge-graph and graph reasoning~\citep{hdreason,HyperGraf,barkam2023reliable}. Our prior work, HyperX~\citep{HyperX}, accelerates Nyström-HDC graph classification on an FPGA, but requires each query graph to fit on chip and executes each hop stage by stage.

\paragraph{HDC accelerators.}
Existing HDC accelerators on FPGAs~\citep{Salamat2019F5HD,FACH,E3HDC}, GPUs~\citep{Kang2022XcelHD}, and processing-in-memory architectures~\citep{hdnn-pim,paap-hd} target applications such as biosignal processing, image classification, and activity recognition~\citep{static-encoding-2-seizure-det,static-encoding-9-hdc-framework-image-descriptors-cvpr,DistriHD,static-encoding-7-efficient-human-activity-recognition-hdc-imani}, accelerating random-projection or record-based encoders over fixed-size inputs. \sys{} instead accelerates HDC for graph-level tasks, where the encoder must process irregular, variable-size graphs.

\paragraph{Graph learning accelerators.}
Most GNN accelerators target node-level tasks on a single large graph, addressing irregular aggregation through hybrid architectures~\citep{HyGCN}, runtime workload rebalancing~\citep{AWBGCN}, overlay architectures~\citep{GraphAGILE}, and dynamic sparsity exploitation~\citep{Dynasparse}, while CPU--FPGA frameworks scale GNN training and inference to large graphs~\citep{GraphACT,HPGNN,BoostGCN}. FlowGNN~\citep{FlowGNN} uses a dataflow architecture for real-time GNN inference on streams of graphs, including graph-level tasks. These accelerators execute learned GNN layers rather than HDC encoding. Stream-dataflow designs~\citep{streaming_arch} and streaming sparse accelerators~\citep{Serpens} further show the benefit of decoupling off-chip memory access from computation. \sys{} applies this principle to graph classification, where each query graph is classified independently and may itself exceed on-chip memory.

\section{From HyperX to \sys{}}
\label{sec:motivation}

\subsection{HyperX and Its Limitations}
\label{subsec:limitations}

HyperX~\citep{HyperX}, our prior FPGA accelerator for Nyström-HDC graph classification, executes Algorithm~\ref{alg:end-to-end} with six compute engines. The Locality Sensitive Hashing Unit (LSHU) computes the LSH codes of each hop by projecting $\mathbf{F}_x$ onto $\mathbf{u}^{(t)}$ and propagating the projection through $\mathbf{A}_x$. The Minimal Perfect Hashing Engine (MPHE) maps each code to its histogram index in $O(1)$ time. The Histogram Update Engine (HUE) accumulates these indices into the hop histogram $\mathbf{h}^{(t)}$, and the Kernel Similarity Engine (KSE) computes $\mathbf{v}^{(t)}=\mathbf{H}^{(t)}\mathbf{h}^{(t)}$ and accumulates it into $\mathbf{C}$. Finally, the Nyström Encoding Engine (NEE) streams $\mathbf{P}_{\mathrm{nys}}$ from DDR to compute the query HV, and the Similarity \& Classification Engine (SCE) matches it against the class prototypes. Combined with DPP-based landmark selection and static load balancing, HyperX achieves low-latency, energy-efficient inference on standard graph classification benchmarks, the largest of which (D\&D) averages 284 nodes and 716 edges per graph. Two design decisions, however, limit HyperX as graphs grow.

\paragraph{On-chip graph storage.}
HyperX buffers the query graph on chip: the feature matrix $\mathbf{F}_x$ and the CSR arrays of $\mathbf{A}_x$ are held in banked BRAM so that the LSHU processing elements (PEs) can access them in parallel. As Table~\ref{tab:mem-complexity} shows, this storage grows with $Nf$ and $\mathrm{nnz}(\mathbf{A}_x)$, whereas the on-chip memory of an edge FPGA is fixed (about 4.5\,MB on the ZCU104) and is shared with the codebooks, hash tables, landmark histograms, and intermediate buffers. The largest graph HyperX can process is therefore bounded by on-chip memory rather than by the application. Graphs in domains such as malware analysis~\citep{Malnet} and robotic object recognition~\citep{neumann2013graph} contain thousands of nodes; MalNet-Tiny graphs, for example, average about 1,400 nodes and 2,900 edges, roughly $5\times$ and $4\times$ those of D\&D.

\paragraph{Stage-sequential execution within each hop.}
Landmark similarity $\mathbf{v}^{(t)}=\mathbf{H}^{(t)}\mathbf{h}^{(t)}$ consumes the complete hop histogram. HyperX therefore processes each hop in stages: the HUE accumulates all indices of the hop and merges its PE-local histograms, and only then does the KSE begin its sparse matrix--vector product. This synchronization barrier recurs at every hop. The KSE idles while the histogram is built, and the LSHU, MPHE, and HUE idle while the KSE computes similarities. Since these engines lie on the critical path of every hop, the barrier increases inference latency and lowers hardware utilization, and its cost grows with the number of nodes processed per hop.

\subsection{Design Approach of \sys{}}
\label{subsec:design-approach}

\sys{} retains the algorithm of HyperX, together with its MPHE, NEE, SCE, and static load balancing, and redesigns the graph-processing front end around three principles.

\paragraph{Stream the query graph from off-chip memory.}
The LSH codes of hop $t$ depend on the graph only through $\mathbf{A}_x^{t}\mathbf{F}_x\mathbf{u}^{(t)}$, so code generation never needs $\mathbf{F}_x$ or $\mathbf{A}_x$ on chip in full. \sys{} replaces the LSHU with a Graph Hashing Unit (GHU) that streams $\mathbf{F}_x$ in row blocks and the CSR arrays of $\mathbf{A}_x$ from DDR through fixed-depth FIFOs, keeping only $O(N)$ state on chip (Section~\ref{sec:ghu}). On-chip storage thus becomes independent of $f$ and $\mathrm{nnz}(\mathbf{A}_x)$, at the cost of re-reading $\mathbf{A}_x$ from DDR in each of the $H(H-1)/2$ propagation steps.

\paragraph{Remove the per-hop barrier.}
Since $\mathbf{H}^{(t)}\mathbf{h}^{(t)}=\sum_j h^{(t)}_j\,\mathbf{H}^{(t)}_{:,j}$, landmark similarity can be accumulated one histogram index at a time, without materializing $\mathbf{h}^{(t)}$. \sys{} fuses the HUE and KSE into a single Histogram Similarity Engine (HSE), which adds column $\mathbf{H}^{(t)}_{:,j}$ to $\mathbf{C}$ for each index $j$ produced by the MPHE (Section~\ref{sec:hse}). Without the barrier, FIFOs connect the GHU, MPHE, and HSE into a streaming dataflow pipeline that overlaps the engines within a hop and lets the GHU begin hop $t+1$ while the MPHE and HSE complete hop $t$ (Section~\ref{subsec:dataflow}).

\paragraph{Retain the benefit of histogram aggregation.}
A complete histogram has one advantage: it aggregates repeated indices before similarity computation, so each distinct index contributes once. Index-by-index accumulation instead accesses column $j$ of $\mathbf{H}^{(t)}$ for every occurrence of index $j$. \sys{} reduces these redundant accesses with window accumulation, which merges repeated indices within a small window into index--count pairs $(j,q)$ and applies $\mathbf{C}\leftarrow\mathbf{C}+q\,\mathbf{H}^{(t)}_{:,j}$, without reintroducing the barrier.

Together, these changes extend Nyström-HDC acceleration to graphs that exceed on-chip memory and reduce inference latency through pipelined execution. Section~\ref{sec:method} details the resulting architecture.

\section{Accelerator Design}
\label{sec:method}

\subsection{Architecture Overview}
\label{subsec:ovw}

\begin{figure}
  \centering
  \includegraphics[width=0.45\columnwidth]{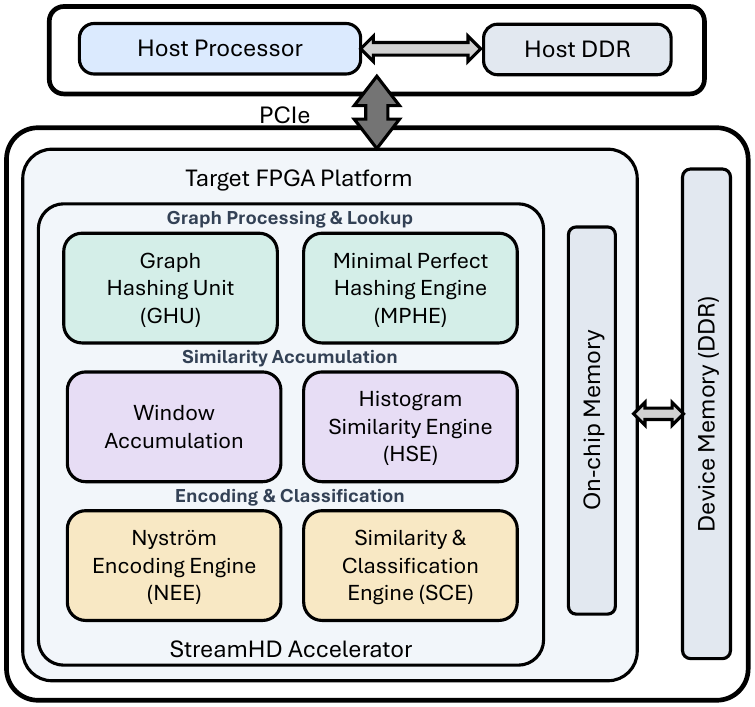}
  \caption{Overview of the \sys{} FPGA accelerator.}
  \label{fig:streamhd-overview}
\end{figure}

Figure~\ref{fig:streamhd-overview} shows the overview of \sys{}, our FPGA accelerator for Nyström-HDC graph classification. \sys{} is composed of five compute engines: (i) the Graph Hashing Unit (GHU), which generates integer codes for nodes at each hop; (ii) the Minimal Perfect Hashing Engine (MPHE), which maps integer codes to valid histogram indices in $O(1)$ time using a minimal perfect hash (MPH); (iii) the Histogram Similarity Engine (HSE), which uses the histogram indices from MPHE to compute hop-level landmark similarities and accumulate the kernel similarity vector $\mathbf{C}$; (iv) the Nyström Encoding Engine (NEE), which projects the kernel vector into the hyperdimensional space; and (v) the Similarity \& Classification Engine (SCE), which computes final class scores by comparing the encoded hypervector against class prototypes, followed by argmax selection of the label. A Window Accumulation Unit between the MPHE and HSE merges repeated histogram indices before similarity accumulation (Section~\ref{subsec:dataflow}).

\subsection{Streaming Dataflow Architecture}
\label{subsec:dataflow}

\begin{figure}
  \centering
  \includegraphics[width=\linewidth]{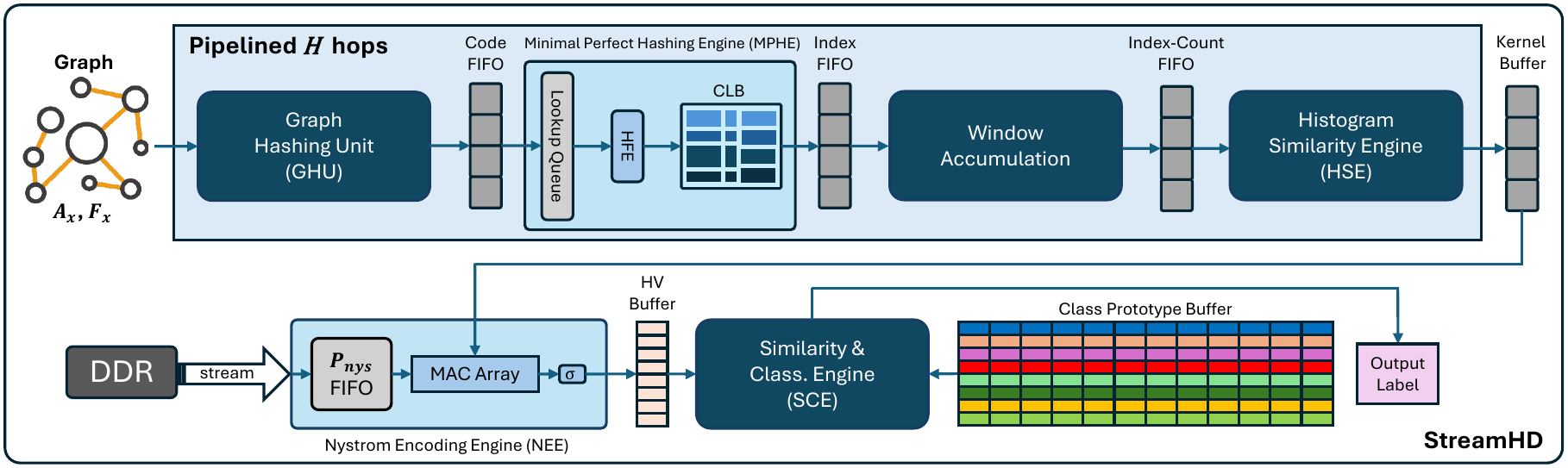}
  \caption{\sys streaming architecture. Code and index FIFOs connect GHU and MPHE to the Window Accumulation Unit, which emits index--count pairs through a FIFO to HSE. HSE accumulates the kernel similarity vector across hops in the Kernel Buffer; NEE and SCE then perform encoding and classification.}
  \label{fig:streamhd-streaming-architecture}
\end{figure}

\begin{figure}
  \centering
  \includegraphics[width=0.75\linewidth]{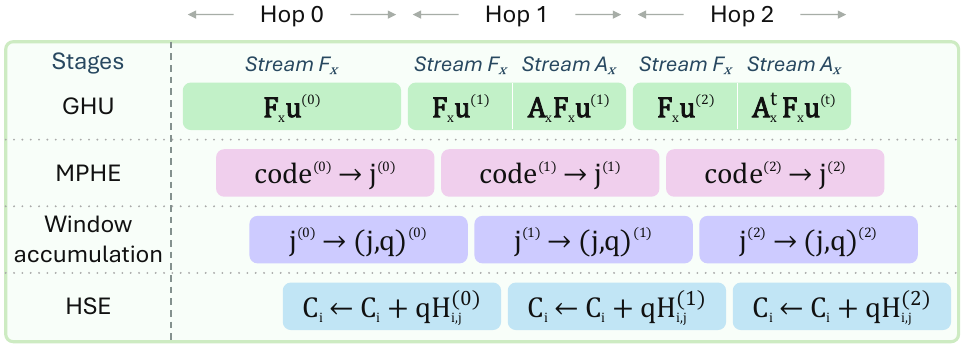}
  \caption{Schematic intra-hop and inter-hop pipelining across GHU, MPHE, the Window Accumulation Unit, and HSE. The Window Accumulation Unit combines histogram indices $j$ into index--count pairs $(j,q)$ for HSE updates. Idle intervals are omitted, and block widths do not represent execution latency. NEE and SCE are not shown.}
  \label{fig:streaming-schedule}
\end{figure}

\begin{algorithm}[t]
\small
\caption{End-to-End Inference in \sys{}. Each step is annotated with the engine that executes it. In hardware, consecutive engines are connected by FIFOs, so the steps of a hop overlap (intra-hop pipelining), and the GHU begins hop $t+1$ while the MPHE and HSE complete hop $t$ (inter-hop pipelining).}
\label{alg:streamhd-inference}
\begin{algorithmic}[1]
\Require Query graph $(\mathbf{A}_x,\mathbf{F}_x)$ and trained model, as in Algorithm~\ref{alg:end-to-end}; window size $W$
\Ensure Query HV $\mathbf{h}\!\in\!\{-1,+1\}^{d}$ and predicted class $\hat{y}$
\State $\mathbf{C}\!\gets\!\mathbf{0}\in\mathbb{R}^{s}$ \Comment{kernel similarity vector}
\For{$t=0$ \textbf{to} $H-1$}
  \State $\mathbf{c}\!\gets\!\mathbf{F}_x\mathbf{u}^{(t)}$ \Comment{GHU: DenseMV over rows of $\mathbf{F}_x$}
  \For{$k=1$ \textbf{to} $t$}
    \State $\mathbf{c}\!\gets\!\mathbf{A}_x\mathbf{c}$ \Comment{GHU: SpMV over CSR of $\mathbf{A}_x$}
  \EndFor
  \State $\mathcal{J}\!\gets\!\emptyset$ \Comment{index window}
  \For{$v=1$ \textbf{to} $N$}
    \State $c_v\!\gets\!\big\lfloor(\mathbf{c}[v]+b^{(t)})/w\big\rfloor$ \Comment{GHU: LSH code}
    \If{$c_v\in\mathcal{B}^{(t)}$}
      \State append $\textproc{Index}(\mathcal{B}^{(t)},c_v)$ to $\mathcal{J}$ \Comment{MPHE}
    \EndIf
    \If{$|\mathcal{J}|=W$ \textbf{or} $v=N$}
      \For{each distinct index $j\in\mathcal{J}$ with count $q$} \Comment{window accumulation}
        \State $\mathbf{C}\!\gets\!\mathbf{C}+q\,\mathbf{H}^{(t)}_{:,j}$ \Comment{HSE}
      \EndFor
      \State $\mathcal{J}\!\gets\!\emptyset$
    \EndIf
  \EndFor
\EndFor
\State $\mathbf{h}\!\gets\!\mathrm{sign}(\mathbf{P}_{\mathrm{nys}}\mathbf{C})$ \Comment{NEE}
\State $\hat{y}\!\gets\!\arg\max_{c}\mathrm{sim}(\mathbf{h},\mathbf{g}_c)$ \Comment{SCE}
\State \Return $\mathbf{h}$ and $\hat{y}$
\end{algorithmic}
\end{algorithm}

Nyström-based graph classification involves several dependent compute stages, but sequential execution leaves downstream PEs idle while waiting for upstream results. This increases inference latency and reduces hardware utilization, which is particularly undesirable for power-constrained edge devices. To address this, we develop a FIFO-based streaming dataflow architecture for \sys{}, as illustrated in Figure~\ref{fig:streamhd-streaming-architecture}, enabling overlapping execution across compute engines. As shown in Algorithm~\ref{alg:streamhd-inference}, \sys{} restructures the inference computation in Algorithm~\ref{alg:end-to-end} to enable pipelined execution across compute engines. Hop-wise propagation uses successive matrix--vector products, reducing intermediate storage from $O(Nf)$ to $O(N)$. This reduces on-chip memory requirements when processing large graphs streamed from DDR (Section~\ref{sec:ghu}). Landmark similarities are accumulated directly into $\mathbf{C}$ using histogram indices, eliminating the histogram-completion barrier and allowing similarity computation to overlap with index generation (Section~\ref{sec:hse}).

The GHU computes the hop-specific vector row by row and forwards each completed row to the Code FIFO for MPHE processing. The MPHE reads these codes through its Lookup Queue, performs codebook lookup, and writes valid histogram indices to the Index FIFO. These FIFO connections decouple code generation from histogram-index lookup, allowing MPHE to process node codes before the GHU completes the current hop.

Each incoming histogram index requires accessing the corresponding column of $\mathbf{H}^{(t)}$. Repeated indices can therefore trigger multiple accesses to the same sparse column, increasing memory traffic. Constructing a complete histogram would eliminate these redundant accesses but reintroduce the hop-level synchronization barrier. To address this, the Window Accumulation Unit combines repeated indices within a fixed-size window into index--count pairs $(j,q)$, where $q$ denotes the occurrence count of index $j$. These pairs are forwarded through the Index-Count FIFO, reducing redundant column accesses without waiting for the complete histogram.

The HSE reads index--count pairs from the Index-Count FIFO and broadcasts them to its PEs. For each pair $(j,q)$, each PE accesses the corresponding sparse column entries of $\mathbf{H}^{(t)}$ within its assigned rows and updates its local similarity buffer as $C_i \leftarrow C_i + qH^{(t)}_{i,j}$. After the final hop, the HSE constructs $\mathbf{C}$ from the PE-local buffers and stores it in the Kernel Buffer. The NEE then generates the hypervector, which is processed by the SCE to produce the classification result.

Figure~\ref{fig:streaming-schedule} illustrates intra-hop and inter-hop pipelining across compute engines. Within hop $t$, the GHU forwards completed elements of $\mathbf{A}_x^t\mathbf{F}_x\mathbf{u}^{(t)}$ while MPHE maps the resulting codes to histogram indices $j^{(t)}$. The Window Accumulation Unit generates index--count pairs $(j,q)^{(t)}$, allowing HSE to begin similarity accumulation before the entire hop completes. Across hops, the GHU begins computing $\mathbf{F}_x\mathbf{u}^{(t+1)}$ while HSE continues updating $\mathbf{C}$ using $\mathbf{H}^{(t)}$. 

\subsection{Compute Engines for Large-Graph Support and Fused Similarity Computation}

\subsubsection{Graph Hashing Unit (GHU)}
\label{sec:ghu}

\begin{figure}
  \centering
  \includegraphics[width=\linewidth]{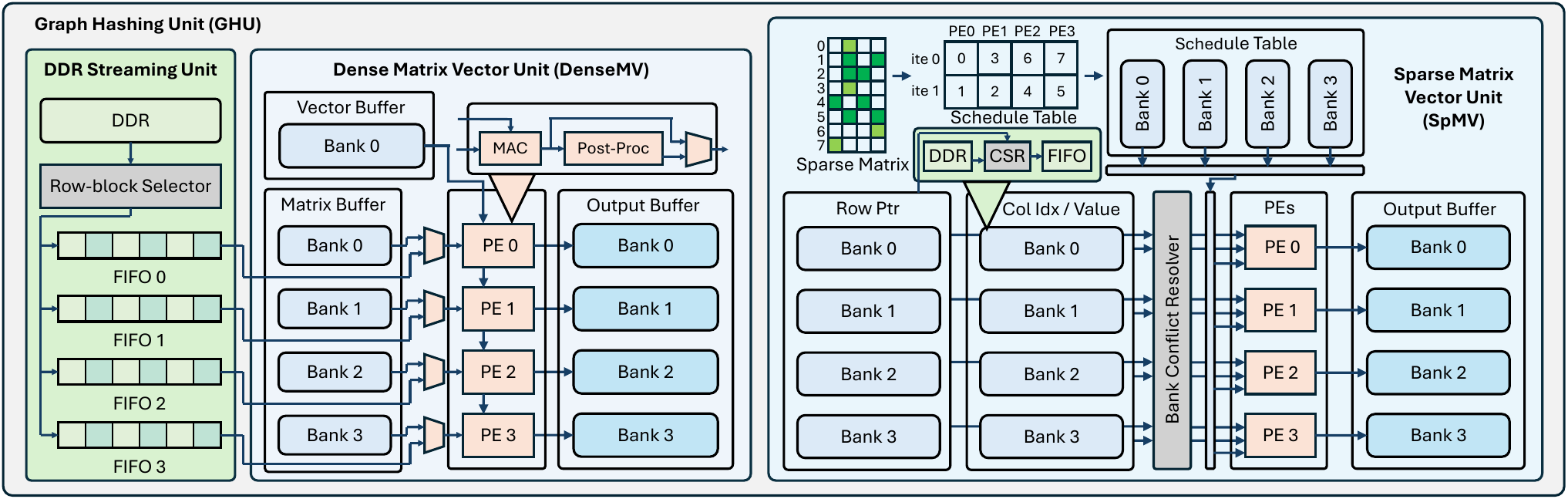}
  \caption{Graph Hashing Unit (GHU) with on-chip buffering for small graphs and DDR streaming for large graphs.}
  \label{fig:ghu-architecture}
\end{figure}

For hop $t$, the GHU first computes the feature projection $\mathbf{c}=\mathbf{F}_x\mathbf{u}^{(t)}$, where $\mathbf{F}_x\in\mathbb{R}^{N\times f}$, $\mathbf{u}^{(t)}\in\mathbb{R}^{f}$, and $\mathbf{c}\in\mathbb{R}^{N}$. Hop-wise propagation then computes $\mathbf{c}_{\mathrm{next}}=\mathbf{A}_x\mathbf{c}$, where $\mathbf{A}_x\in\{0,1\}^{N\times N}$ and $\mathbf{c}_{\mathrm{next}}\in\mathbb{R}^{N}$, followed by $\mathbf{c}\leftarrow\mathbf{c}_{\mathrm{next}}$. Repeating this operation $t$ times computes the required $\mathbf{A}_x^{t}\mathbf{F}_x\mathbf{u}^{(t)}$. For graphs with large numbers of nodes and edges, the storage requirements of the feature matrix $\mathbf{F}_x$ and adjacency matrix $\mathbf{A}_x$ can exceed the on-chip memory capacity of edge FPGAs. To address this, the GHU supports two memory configurations (Figure~\ref{fig:ghu-architecture}): on-chip buffering for graphs with a small number of nodes and DDR streaming of $\mathbf{F}_x$ and $\mathbf{A}_x$ for large graphs.

The dense matrix--vector (DenseMV) unit computes the feature projection across multiple PEs. For small graphs, $\mathbf{F}_x$ is stored in an on-chip matrix buffer banked across PEs. For large graphs, the DDR streaming unit reads $\mathbf{F}_x$ in contiguous row blocks of size $N_b \times f$, requiring $\lceil N/N_b \rceil$ blocks. The row-block selector cyclically distributes feature rows across the PE input FIFOs. These FIFOs enable overlapping DDR reads with PE computation while maintaining bounded input buffering. The projection vector $\mathbf{u}^{(t)}$ is stored in registers to support concurrent reads from PEs. Each PE computes the dot product of its assigned feature rows with $\mathbf{u}^{(t)}$ using a dedicated MAC unit and writes the results to the banked output vector buffer. 

The sparse matrix--vector (SpMV) unit performs hop-wise propagation using the CSR representation of $\mathbf{A}_x$. The CSR row pointers are stored in banked on-chip memory, while the column indices and nonzero values are buffered on-chip for small graphs or streamed from DDR for large graphs. For large graphs, the CSR streaming unit uses the row pointers to identify the nonzero entries of each row and reads them sequentially from DDR into the PE input FIFOs. Each FIFO entry contains the row index, column index, nonzero value, and a flag indicating the last nonzero of the row. The column indices determine the required elements of the banked input vector $\mathbf{c}$, while the nonzero values are used for sparse multiply-accumulate operations.

Irregular sparsity in $\mathbf{A}_x$ can cause load imbalance across PEs. To avoid runtime scheduling overhead, we use a static iteration-wise load balancer that assigns rows to PEs offline based on their nonzero counts. The schedule table is constructed by (1) computing $\texttt{nnz}[r]$ for each row; (2) grouping rows into buckets based on $\texttt{nnz}$; and (3) assigning $P$ rows with similar nonzero counts to each iteration. Given $N$ rows and $P$ PEs, the resulting $\lceil N/P\rceil \times P$ schedule table provides the row assigned to each PE in every iteration. The table is stored in BRAM and banked along columns, allowing PEs to read their assigned rows concurrently without runtime scheduling.

In SpMV, each PE accesses the banked intermediate vector $\mathbf{c}$ using the column indices of its assigned CSR row. Multiple PEs may request elements from the same memory bank, exceeding the available read ports. The Bank Conflict Resolver arbitrates these requests using a round-robin policy, granting up to two reads per bank and deferring the remaining requests. The retrieved vector elements are multiplied with the corresponding nonzero values and accumulated locally, with the completed row results written to the banked output vector $\mathbf{c}_{\mathrm{next}}$. Once each node's hop-specific value $c_i$ is computed, it is quantized into an integer LSH code $\lfloor(c_i+b^{(t)})/w\rfloor$, where $b^{(t)}$ and $w$ denote the hop-specific offset and quantization width, respectively.

In summary, the GHU scales to larger graphs by streaming feature and adjacency data through fixed-depth FIFOs, avoiding on-chip storage of the complete matrices while retaining the intermediate vectors on-chip.

\subsubsection{Histogram Similarity Engine (HSE)}
\label{sec:hse}

\begin{figure}
  \centering
  \includegraphics[width=0.8\linewidth]{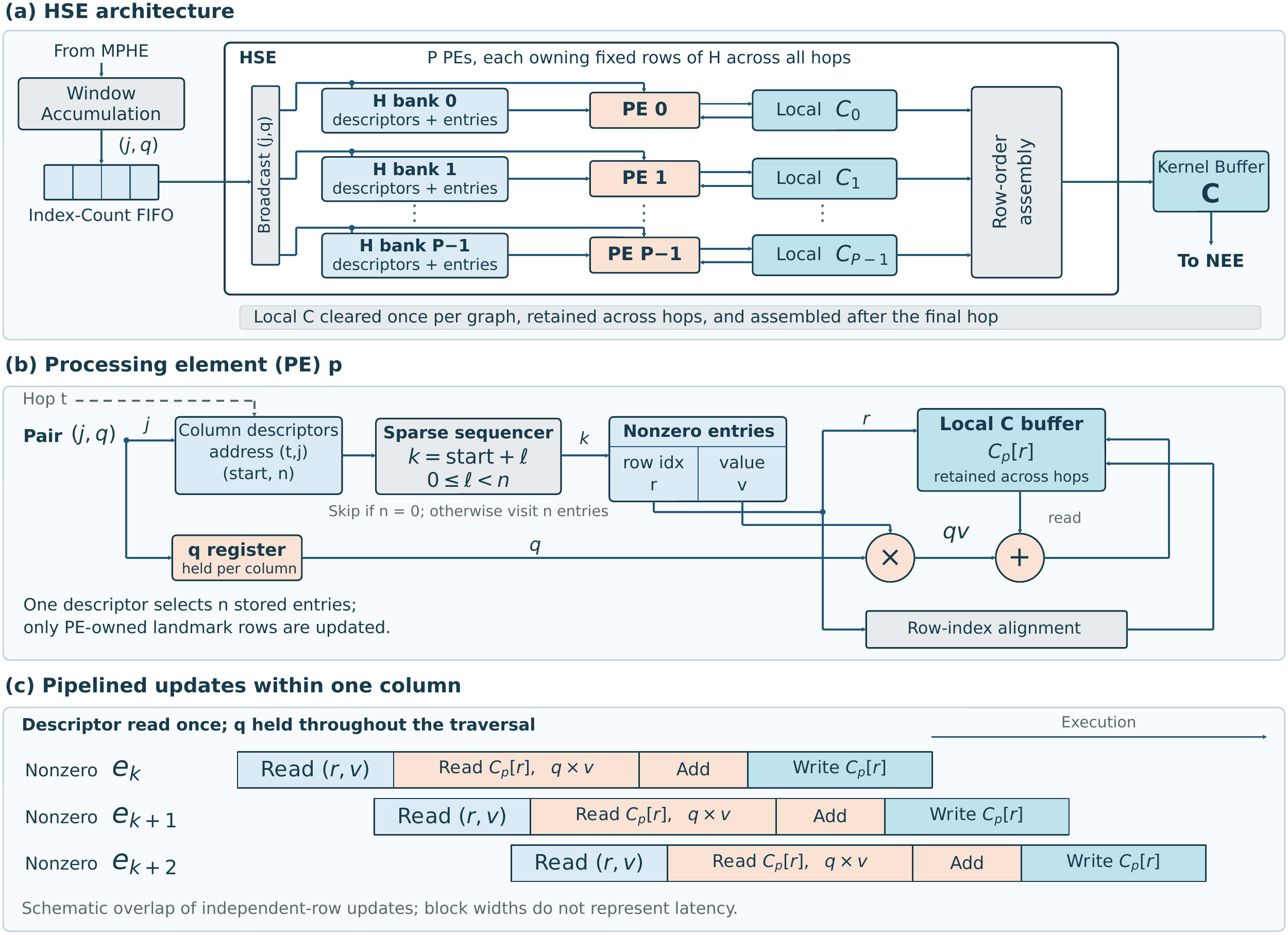}
  \caption{Histogram Similarity Engine (HSE). (a) Index--count pairs $(j,q)$ from the Window Accumulation Unit are broadcast to $P$ PEs, each owning a fixed set of rows of $\mathbf{H}^{(t)}$. (b) Each PE reads the descriptor $(\mathrm{start},n)$ of column $j$ and, for each stored entry $(r,v)$ at address $k$, accumulates $q\,v$ into its local $\mathbf{C}$ buffer. (c) Updates for consecutive entries $e_k, e_{k+1}, \dots$ of a column belong to distinct rows and are therefore pipelined.}
  \label{fig:hse-architecture}
\end{figure}

The HSE computes hop-wise landmark similarities $\mathbf{v}^{(t)}$ and accumulates them into the kernel similarity vector $\mathbf{C}$. In the conventional formulation, for each hop $t$, the histogram indices are first accumulated into the query histogram $\mathbf{h}^{(t)}$, followed by similarity computation $\mathbf{v}^{(t)}=\mathbf{H}^{(t)}\mathbf{h}^{(t)}$. Since similarity computation requires the completed histogram, this formulation introduces a hop-level synchronization barrier that prevents the two operations from overlapping.

To eliminate this barrier, the HSE directly accumulates landmark similarities from incoming histogram indices. For each index $j$ at hop $t$, the corresponding column of $\mathbf{H}^{(t)}$ is added to the kernel similarity vector as $\mathbf{C}\leftarrow\mathbf{C}+\mathbf{H}^{(t)}_{:,j}$. This eliminates the need to construct the complete query histogram before similarity computation. The Window Accumulation Unit (Section~\ref{subsec:dataflow}) combines repeated indices into $(j,q)$ pairs, where $q$ denotes the occurrence count. The HSE then updates $\mathbf{C}\leftarrow\mathbf{C}+q\mathbf{H}^{(t)}_{:,j}$. This replaces repeated accesses to the same sparse column within a window with a single column lookup and weighted accumulation, reducing memory traffic and computation.

As shown in Figure~\ref{fig:hse-architecture}(a), the HSE partitions the rows of $\mathbf{H}^{(t)}$ across PEs, with each PE owning a disjoint set of rows across all hops. Each PE maintains the corresponding entries of $\mathbf{C}$ in a local buffer. To support index-based access, the assigned rows are organized in a column-oriented sparse format. Each column descriptor specifies the starting offset and number of stored entries, with each entry containing a local row index and its corresponding value. Each incoming index--count pair $(j,q)$ is broadcast to all PEs. For hop $t$, each PE accesses the descriptor for column $j$, retrieves its nonzero entries, and accumulates their count-weighted values into its local $\mathbf{C}$ buffer (Figure~\ref{fig:hse-architecture}(b)). Since the entries of a column belong to distinct rows, successive updates within a column are independent and are pipelined (Figure~\ref{fig:hse-architecture}(c)). Since $\mathbf{H}^{(t)}$ is sparse, this avoids processing zero entries and reduces memory accesses and computation. The local buffers retain accumulated similarities across hops. After the final hop, the HSE constructs $\mathbf{C}$ from the PE-local buffers according to the assigned row indices.

\subsection{Additional Compute Engines}
This section briefly summarizes the MPHE, NEE, and SCE introduced in our prior work~\cite{HyperX} to provide a complete description of \sys{}.

\subsubsection{Minimal Perfect Hashing Engine (MPHE)}
\label{subsec:mphe}
We adopt an MPH scheme~\cite{BBHash} to map codes (keys) in the codebook to indices $\{0,\dots,|\mathcal{B}^{(t)}|\!-\!1\}$ with $O(1)$ query time during inference. The scheme constructs the MPH as a cascade of \emph{levels} $A_0, A_1,\dots$, where each bit-array $A_d$ is sized to balance memory against collision probability. During construction, keys hashing uniquely at level $d$ set a $1$ at position $h_d(\text{key})$ and are removed, while colliding keys advance to the next level $d{+}1$. The final structure concatenates all bit-arrays with a lightweight \emph{rank} vector. At query time, a key traverses the levels until it encounters a $1$, and its index is computed as the cumulative number of 1s preceding that position. This yields a compact ($\approx\!3$ bits/key) structure supporting constant-time queries.
The MPHE performs inference-time dictionary lookup from code $\rightarrow$ histogram index. Upstream, the GHU produces a chunk of integer codes, which are enqueued into a \emph{lookup queue}. Each lookup then proceeds as follows:
\begin{enumerate}
    \item \emph{Hash generation.} A \emph{Hash Function Engine} computes two 64-bit hashes for the input code using a seeded integer hash function \cite{integerhash}. The level-$d$ probe index is $i_d \gets h_d(\text{code}) \bmod |A_d|$, where $|A_d|$ is the bit capacity of level table $A_d$. 
    \item \emph{Level table probe.} Level tables $\{A_d\}$ are stored on-chip as banks of 64-bit words; the bit at $i_d$ is checked within its word. If it is 0, the code advances to the next level by generating the next hash in the sequence via a xorshift-based rehash generator \cite{sequencehashgen}. 
    \item \emph{Rank to MPH index.} On a 1-bit hit at $(d,i_d)$, the engine retrieves the global index directly from the rank vector. Each entry in the rank vector stores the cumulative number of 1s up to the start of a word, aggregated across all levels. The final index is obtained by adding this value to the popcount of bits within the current word up to (and including) position $i_d$, and subtracting one for zero-based indexing. 
    \item \emph{Codebook verification.} The MPH index addresses a compact \emph{codebook store} that holds pairs \texttt{(code,hist\_idx)}. The queried code is compared against the stored code. If they match, the corresponding \texttt{hist\_idx} is emitted as valid. If they differ, the lookup terminates, as a hit with a mismatch implies the queried code is not in the original codebook. If all levels are probed with no hit, the code is deemed absent. No histogram update occurs for a queried code absent from the codebook.
\end{enumerate}

The MPHE datapath consists of a \emph{Lookup Queue} that collects codes from the GHU PEs, a \emph{Hash Function Engine} that generates 64-bit hashes and level indices, \emph{Level Tables} $A_d$ and \emph{Rank Vectors} that are banked across BRAMs and accessed in parallel, an \emph{Index Buffer} that carries the computed MPH index, and a banked \emph{Codebook} that returns the \texttt{(code,hist\_idx)} pair to a \emph{Compare Unit} that verifies code presence. Level tables and rank vectors are placed in independent banks to support pipelined parallel lookups, while contention in codebook access is reduced through MPH hashing, pipelining, and banking. The pipeline roughly issues one index lookup per cycle, with $O(1)$ cost bounded by the small number of level tables per codebook.

\subsubsection{Nyström Encoding Engine (NEE)}

The NEE generates the final hypervector through projection $\mathbf{h} = \mathrm{sign}(\mathbf{P}_{\mathrm{nys}} \mathbf{C})$, where $\mathbf{P}_{\mathrm{nys}}$ is the Nyström projection matrix and $\mathbf{C}$ is the kernel similarity vector from the HSE. Storing $\mathbf{P}_{\text{nys}}$ on-chip is infeasible for edge devices with limited on-chip memory; hence, $\mathbf{P}_{\mathrm{nys}}$ is streamed from external DDR during inference. This makes the NEE a performance-critical engine. 
To address this bottleneck, this work adopts a streaming (dataflow)~\cite{streaming_arch} architecture in which compute stages are connected via on-chip FIFOs to overlap data movement with execution. 
To maximize external memory bandwidth utilization, data transfers are aligned to the FPGA platform’s native memory-port width, $y$ bits. Accordingly, we organize $\mathbf{P}_{\mathrm{nys}}$ into $y$-bit words and issue contiguous $y$-bit burst transfers. This maximizes effective bandwidth utilization, avoiding inefficiency from irregular or undersized accesses. We also enable multiple outstanding reads to keep the memory interface saturated despite variable off-chip DRAM latency. Incoming $y$-bit words are buffered in a deep on-chip FIFO, decoupling bursty DRAM traffic from the compute stage. 
Each $y$-bit word is unpacked into $y$/$x$ operands, where $x$ is the precision of the operand in bits, and is routed to $y$/$x$ MAC lanes, fully utilizing the memory interface width. Each MAC lane performs a dot product between its assigned elements of $\mathbf{P}_{\mathrm{nys}}$ and the corresponding entries of the similarity vector $\mathbf{C}$, which is buffered on chip with cyclic partitioning to enable conflict-free parallel access. After accumulation, results are bipolarized to $\{-1,+1\}$.

\subsubsection{Similarity \& Classification Engine (SCE)}
\label{subsec:sce}

The SCE performs the final stage of inference by comparing the encoded hypervector $\mathbf{h}$ against the stored class prototypes $\{\mathbf{g}_c\}_{c=1}^C$. This is implemented as a dense matrix–vector multiplication $\mathbf{s} = \mathbf{G} \mathbf{h}$, where $\mathbf{G} \in \{-1,+1\}^{C \times d}$. The computation is parallelized across PEs, each responsible for a block of rows in $G$. The output scores $\mathbf{s}$ are written into an on-chip buffer, followed by a small argmax unit that identifies the predicted class $\hat{y} = \arg\max_c s_c$. 
\section{Experimental Results}
\label{sec:results}

\subsection{Experimental Setup}
\label{sec:experimental-setup}

\subsubsection{Implementation Details}
\label{sec:implementations-baselines}
We implement \sys on the AMD Zynq UltraScale+ ZCU104 FPGA~\cite{zcu104}, representative of an edge-class FPGA. The design targets real-time inference with batch size~1, consistent with edge deployment scenarios where query graphs arrive sequentially. The accelerator is designed using AMD Vitis High Level Synthesis (HLS) and synthesized using AMD Vitis Unified IDE v2024.2~\cite{AMD_Vitis_Unified_IDE_2024_2}. We configure 2--5 PEs across the GHU, MPHE, and HSE based on their relative computational workloads, which depend on the number of nodes, nonzeros in the adjacency matrix, and propagation hops for each dataset. This balances execution across stages while limiting FPGA resource utilization. For large graphs, the GHU streams feature rows in blocks of $N_b=64$, distributed cyclically across the DenseMV PEs. The Window Accumulation Unit uses four independent 32-index windows, one per MPHE PE. The hypervector dimension is set to $d=10{,}000$ for all datasets except REDDIT-Multi-5K and FIRSTMM\_DB, which use $d=20{,}000$. For NEE, $\mathbf{P}_{\mathrm{nys}}$ is streamed through a 512-bit AXI transfer width. We configure the inter-stage FIFOs with a depth of 128 entries, with widths matched to the corresponding data types of each stage. For the large-graph configurations, two NEE units process alternating rows of $\mathbf{P}_{\mathrm{nys}}$ to increase projection parallelism. Table~\ref{tab:firstmm-resource-utilization} reports the FPGA resource utilization at an achieved frequency of 300\,MHz.


\subsubsection{Datasets}
\label{sec:datasets-protocol}

We evaluate \sys{} on eleven benchmark graph classification datasets spanning bioinformatics, chemistry, cybersecurity, social networks, and computer vision. These include eight benchmarks from TUDataset~\cite{tudatasetcollectionbenchmarkdatasets} and three datasets with larger graphs: MalNet-Tiny, REDDIT-Multi-5K, and FIRSTMM\_DB. In this work, large graphs are those for which the feature and adjacency matrices can exceed the on-chip memory of an edge FPGA, such as MalNet-Tiny and FIRSTMM\_DB. In prior FPGA accelerator work~\cite {HyperX}, D\&D was the largest benchmark by average graph size, with 284 nodes and 716 edges per graph. Compared with D\&D, we evaluate on MalNet-Tiny, which contains approximately $5\times$ more nodes and $4\times$ more edges per graph. Our evaluation focuses on graph-level classification of independent input graphs, rather than node- or link-level prediction on a single massive graph. Table~\ref{tab:datasets} summarizes the dataset statistics.

\begin{table}[ht]
  \centering
  \caption{Summary of graph classification datasets.}
  \label{tab:datasets}
  \resizebox{\columnwidth}{!}{%
  \begin{tabular}{llrrr}
    \toprule
    \textbf{Application} & \textbf{Dataset} & \textbf{\#Graphs}
    & \textbf{Avg. Nodes} & \textbf{Avg. Edges} \\
    \midrule
    Bioinformatics & ENZYMES~\cite{Borgwardt2005}
      & 600   & 33    & 62    \\
    & D\&D~\cite{Dobson2003}
      & 1,178 & 284   & 716   \\
    \midrule
    Small molecule & NCI1~\cite{molecule}
      & 4,110 & 30    & 32    \\
    & NCI109~\cite{molecule}
      & 4,127 & 30    & 32    \\
    & BZR~\cite{Sutherland2003}
      & 405   & 36    & 38    \\
    & COX2~\cite{Sutherland2003}
      & 467   & 41    & 43    \\
    & MUTAG~\cite{Debnath1991}
      & 188   & 18    & 20    \\
    & Mutagenicity~\cite{Riesen2008}
      & 4,337 & 30    & 31    \\
    \midrule
    Cybersecurity & MalNet-Tiny~\cite{Malnet}
      & 5,000 & 1,410 & 2,860 \\
    \midrule
    Social network & REDDIT-Multi-5K~\cite{10.1145/2783258.2783417}
      & 4,999 & 509   & 595   \\
    \midrule
    Computer vision & FIRSTMM\_DB~\cite{neumann2013graph}
      & 41    & 1,377 & 3,074 \\
    \bottomrule
  \end{tabular}}
\end{table}

\subsubsection{Baseline Platforms}
\label{sec:platforms}
We evaluate \sys against CPU and GPU baselines (Table~\ref{tab:baseline-platforms}): the AMD Ryzen 5 5625U and the NVIDIA RTX A4000. All baselines are implemented in \texttt{PyTorch} (v2.4.1) with Python~3.8 and CUDA~12.1. We select the Ryzen processor as representative of CPUs deployed in laptops and compact systems, while the RTX A4000 is a workstation-class GPU commonly used as a baseline in prior HDC works~\cite{hdreason, hdnn-pim, paap-hd, survey-edge}. Both baselines offer substantially higher peak compute and memory resources than our target edge FPGA, providing a conservative comparison. For fairness, GPU latency is measured after all model parameters are transferred to device memory, whereas on the FPGA, the Nyström projection matrix—the dominant model parameter—must be streamed from DDR during inference. Despite this, \sys consistently achieves lower latency. While workstation-class GPUs naturally draw more power, we report energy (mJ/graph) to ensure fairness.

\begin{table}
  \centering
  \caption{Specifications of baseline platforms.}
  \label{tab:baseline-platforms}
  \footnotesize
  \setlength{\tabcolsep}{5pt}
  \begin{tabular}{lccc}
    \toprule
    & \textbf{CPU} & \textbf{GPU} & \textbf{FPGA (Ours)} \\
    \midrule
    Platform
      & AMD Ryzen 5 5625U & NVIDIA RTX A4000 & AMD ZCU104 \\
    Frequency
      & 2.3 GHz (base) & 1560 MHz & 300 MHz \\
    Peak Perf.
      & 2.4 TFLOPS (FP32) & 19.2 TFLOPS (FP32) & 0.26 TFLOPS (FP32) \\
    On-chip Mem.
      & 3 MB L2, 16 MB L3 & 6 MB L2 Cache & 4.5 MB \\
    Memory BW
      & 50 GB/s (DDR4) & 448 GB/s (GDDR6) & 19.2 GB/s (DDR4) \\
    \bottomrule
  \end{tabular}
\end{table}


\subsubsection{Metrics}
\label{sec:metrics-methodology}
We report four key metrics:  
(i) \emph{Classification accuracy} (\%): fraction of correctly predicted labels against ground truth;  
(ii) \emph{Latency} (ms/graph): average end-to-end inference time per graph at batch size 1, measured from the arrival of the graph’s adjacency and feature matrices to the prediction of the class label;
(iii) \emph{Power} (W): average device power during steady-state inference, including static and dynamic components;  
(iv) \emph{Energy} (mJ/graph): average energy per inference, computed as the product of measured power and latency.

\subsection{Results and Analysis}
\label{sec:results-analysis}

\subsubsection{End-to-End Performance Comparison with Baseline Platforms}
\label{sec:performance-energy}

\begin{table*}[width=\textwidth,pos=t]
\caption{CPU, GPU, and FPGA latency, power, and energy per graph comparison.}
\label{tab:system-performance-energy}
\centering
\normalsize
\setlength{\tabcolsep}{4.5pt}
\renewcommand{\arraystretch}{1.12}
\begin{tabular*}{\textwidth}{@{\extracolsep{\fill}}lrrrrrrrrr@{}}
\toprule
& \multicolumn{3}{c}{Latency (ms/graph)}
& \multicolumn{3}{c}{Power (W)}
& \multicolumn{3}{c}{Energy (mJ/graph)} \\
\cmidrule(lr){2-4}\cmidrule(lr){5-7}\cmidrule(l){8-10}
Dataset & CPU & GPU & FPGA & CPU & GPU & FPGA & CPU & GPU & FPGA \\
\midrule
D\&D & 6.11 & 3.00 & 0.49 & 25.20 & 59.00 & 1.94 & 154.00 & 177.00 & 0.96 \\
ENZYMES & 2.55 & 1.60 & 0.24 & 24.30 & 61.00 & 1.76 & 62.00 & 98.00 & 0.43 \\
MUTAG & 3.87 & 4.90 & 0.27 & 25.30 & 60.00 & 1.98 & 98.00 & 294.00 & 0.54 \\
NCI1 & 4.23 & 2.60 & 0.26 & 25.50 & 61.00 & 2.05 & 108.00 & 159.00 & 0.53 \\
BZR & 2.29 & 1.60 & 0.24 & 24.60 & 62.00 & 1.56 & 56.00 & 99.00 & 0.37 \\
COX2 & 4.68 & 6.70 & 0.26 & 24.40 & 61.00 & 1.93 & 114.00 & 409.00 & 0.50 \\
NCI109 & 3.44 & 2.60 & 0.27 & 24.90 & 61.00 & 1.83 & 86.00 & 159.00 & 0.48 \\
Mutagenicity & 3.01 & 1.70 & 0.25 & 24.40 & 60.00 & 1.88 & 73.00 & 102.00 & 0.47 \\
MalNet-Tiny & 22.39 & 0.75 & 0.55 & 27.10 & 84.00 & 1.68 & 606.77 & 63.20 & 0.92 \\
REDDIT-Multi-5K & 6.81 & 1.18 & 1.11 & 25.40 & 65.00 & 1.99 & 172.97 & 76.80 & 2.21 \\
FIRSTMM\_DB & 56.17 & 7.04 & 11.39 & 26.30 & 67.00 & 2.35 & 1477.27 & 471.42 & 26.71 \\
\bottomrule
\end{tabular*}
\end{table*}

We evaluate the end-to-end inference latency of \sys{} against CPU and GPU baselines at batch size one under identical algorithmic settings. For FPGA, latency is measured using the integrated analyzer in AMD Vitis Unified IDE v2024.2 at 300\,MHz. CPU and GPU latencies are measured using wall-clock timing in PyTorch. Table~\ref{tab:system-performance-energy} summarizes the results across all eleven datasets.

Across all datasets, \sys{} achieves average latency speedups of $14.3\times$ and $8.4\times$ over CPU and GPU, respectively. These improvements are due to FPGA-specific datapaths that exploit PE-level parallelism, pipelined execution, and efficient on-chip buffering to reduce latency and overlap computation with data movement. Unlike throughput-oriented GPUs, \sys{}'s fine-grained datapath architecture is custom-designed for batch-size-one inference.
On FIRSTMM\_DB, \sys achieves a latency of 11.39\,ms on FPGA compared with 7.04\,ms on GPU. Despite having an average graph size comparable to MalNet-Tiny, FIRSTMM\_DB requires 16 hops, increasing repeated sparse matrix computations and memory accesses. This illustrates that inference latency depends not only on graph size but also on propagation depth, adjacency sparsity, landmark count, and hypervector dimension.

CPU power is measured using a PM01-US plug-in power meter, GPU power using \texttt{nvidia-smi}~\cite{nvidiasmi}, and FPGA power using post-implementation reports in AMD Vitis Unified IDE~\cite{AMD_Vitis_Unified_IDE_2024_2}, consistent with standard FPGA power estimation~\cite{Salamat2019F5HD}. As shown in Table~\ref{tab:system-performance-energy}, \sys{} consumes 1.56--2.35\,W, compared with 24.3--27.1\,W on CPU and 59--84\,W on GPU. Across all datasets, \sys{} achieves average energy reductions of $199.7\times$ and $273.8\times$ over CPU and GPU, respectively. These improvements arise from the proposed architecture, which reduces computation and data movement through pipelined execution and efficient on-chip buffering.

\subsubsection{Impact of Streaming Architecture}
\label{sec:architectural-ablation}

\begin{figure*}
  \centering
  \includegraphics[width=\textwidth]{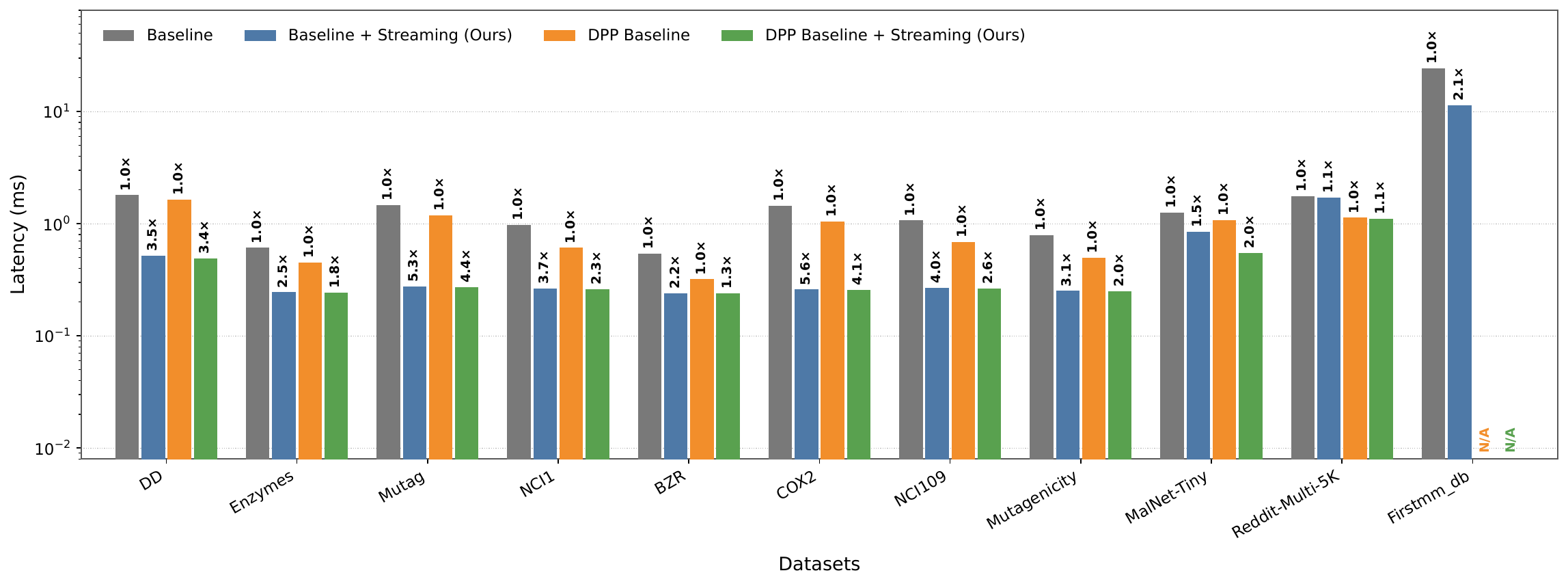}
  \caption{FPGA latency comparison between Baseline, Baseline + Streaming, DPP Baseline, and DPP Baseline + Streaming configurations. Labels show speedup relative to the corresponding non-streaming baseline of each pair; both baselines are normalized to $1.0\times$.}
  \label{fig:streaming-comparison}
\end{figure*}

Figure~\ref{fig:streaming-comparison} evaluates the impact of the proposed streaming architecture on FPGA inference latency using four configurations: Baseline, Baseline + Streaming, DPP Baseline, and DPP Baseline + Streaming. The Baseline implements the Nyström-HDC algorithm from NysHD~\cite{NysHD}, while DPP Baseline uses the DPP-based landmark reduction from prior work~\cite{HyperX} to reduce the number of landmarks and inference latency. The two streaming configurations include our proposed architecture along with their respective algorithmic baselines, isolating the impact of streaming with and without DPP. DPP-based landmark reduction is not applied to FIRSTMM\_DB because the dataset contains only 41 graphs.

Across all datasets, streaming achieves an average speedup of $3.1\times$ without DPP, while the DPP + Streaming configurations achieve an average speedup of $2.5\times$ across ten datasets, relative to their respective non-streaming baselines. The highest speedup of $5.6\times$ is observed on COX2 without DPP. On MalNet-Tiny, streaming achieves $1.5\times$ and $2.0\times$ speedups without and with DPP, respectively. These improvements result from FIFO-based dataflow and fused histogram similarity computation, which reduce synchronization overhead and enable concurrent execution across compute engines.

The benefits of streaming also depend on the propagation hops. FIRSTMM\_DB, with 16 hops, achieves a $2.1\times$ speedup, providing opportunities for both intra-hop and inter-hop pipelining. In contrast, REDDIT-Multi-5K uses a single hop and shows limited improvement, as inter-hop pipelining cannot be exploited. These results demonstrate the effectiveness of \sys{}'s streaming architecture, particularly for workloads with many propagation hops.

\subsubsection{Classification Accuracy}
\label{sec:classification-accuracy}

\begin{figure*}
  \centering
  \includegraphics[width=\textwidth]{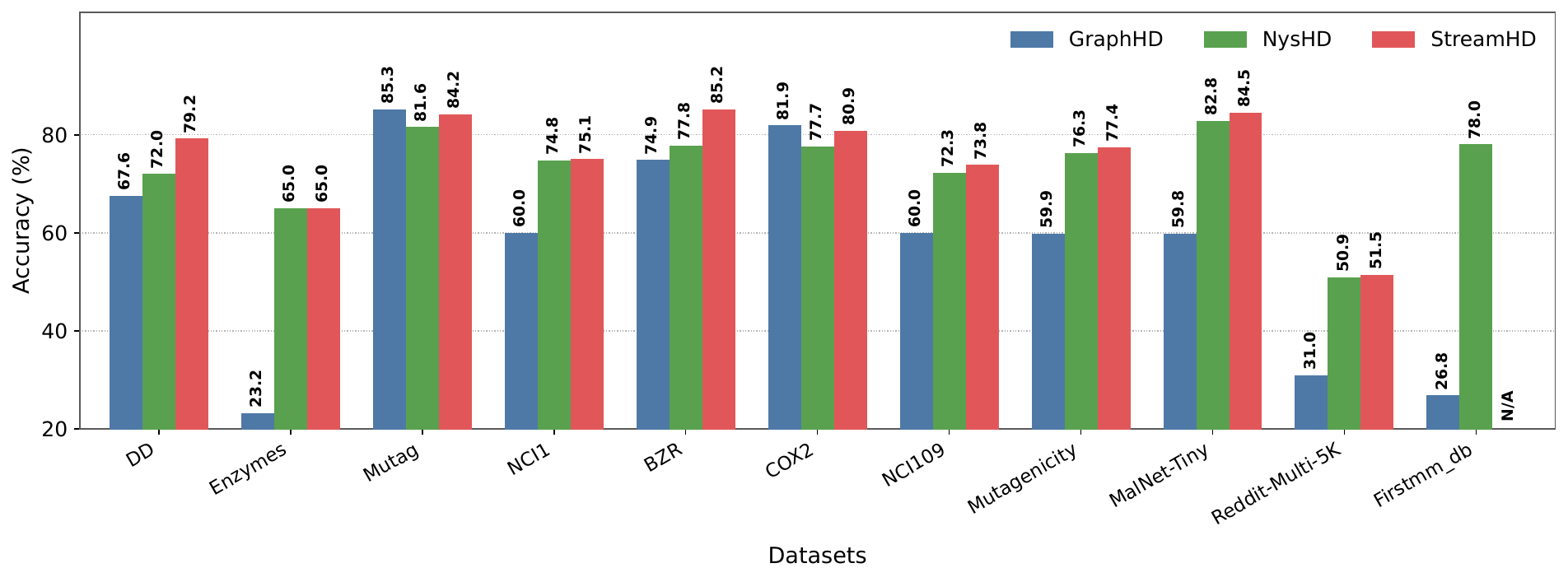}
  \caption{Classification accuracy (\%) on graph datasets.}
  \label{fig:accuracy-comparison}
\end{figure*}

We evaluate classification accuracy on the three additional datasets to assess the performance of existing HDC-based graph classification methods on larger graph workloads. Figure~\ref{fig:accuracy-comparison} compares GraphHD~\cite{GrapHD}, NysHD~\cite{NysHD}, and \sys across the eleven datasets. For the three additional datasets, we reimplement the methods under consistent experimental settings. On MalNet-Tiny, \sys achieves $84.5\%$ accuracy, compared to $82.8\%$ in NysHD and $59.8\%$ in GraphHD. Similarly, on REDDIT-Multi-5K, \sys achieves $51.5\%$, compared with $50.9\%$ and $31.0\%$, respectively. On FIRSTMM\_DB, NysHD achieves $78.0\%$ accuracy, substantially outperforming GraphHD at $26.8\%$. 
These results demonstrate the effectiveness of Nyström-based HDC methods on graph classification.

\subsubsection{Resource Utilization and Timing}
\label{sec:resources-timing}

Table~\ref{tab:firstmm-resource-utilization} reports the resource utilization obtained using the Vitis Unified IDE, with an achieved frequency of \(300\,\text{MHz}\). This high clock rate is enabled by compact architectural components with localized datapath and BRAM-banked on-chip memory, reducing critical-path length. We map the on-chip storage to BRAM rather than URAM as it provides lower access latency, flexible bit width, and higher effective bandwidth, which are critical for edge applications. With a maximum BRAM utilization of 62\%, the results demonstrate that the largest evaluated graph configuration fits within the target FPGA's resource budget.

\begin{table}
\caption{Resource utilization of the FIRSTMM\_DB dataset design on ZCU104.}
\label{tab:firstmm-resource-utilization}
\centering
\small
\setlength{\tabcolsep}{6pt}
\renewcommand{\arraystretch}{1.08}
\begin{tabular}{lrrr}
\toprule
Resource & Used & Available & Utilization \\
\midrule
LUT        & 78,517 & 230,400 & 34\% \\
FF         & 70,988 & 460,800 & 15\% \\
BRAM (18K) & 385    & 624     & 62\% \\
DSP        & 119    & 1,728   & 7\% \\
URAM       & 0      & 96      & 0\% \\
\bottomrule
\end{tabular}
\end{table}


\section{Conclusion}
\label{sec:conclusion}

We presented \sys{}, an FPGA accelerator for low-latency and energy-efficient Nyström-HDC graph classification on edge devices. \sys{} streams the feature and adjacency data of large graphs from DDR, supporting graphs that exceed on-chip memory. We introduced a streaming dataflow architecture that removes stage-wise synchronization barriers, enables pipelined execution, and achieves an average $3.1\times$ speedup over the non-streaming implementation. Compared with CPU and GPU baselines, \sys{} achieves average latency speedups of $14.3\times$ and $8.4\times$, respectively, while reducing energy per graph by $199.7\times$ and $273.8\times$.

\section*{Declaration of Generative AI and AI-Assisted Technologies in the Manuscript Preparation Process}

During the preparation of this work, the authors used ChatGPT/Claude in order to improve the presentation of this manuscript. After using this tool/service, the authors reviewed and edited the content as needed and take full responsibility for the content of the published article.


\clearpage
\bibliographystyle{cas-model2-names}

\bibliography{references}


\end{document}